\documentclass[%
	epj
  ]{svjour}[1999/09/14]
\usepackage{amsmath}
\usepackage{amssymb}
\usepackage{mathrsfs}
\usepackage[sectionbib,numbers,sort&compress]{natbib}
\usepackage{graphicx}
\usepackage{color}
\ifx\pdfoutput\undefined
\DeclareGraphicsExtensions{.mps,.ps,.eps}
\providecommand{\eprint}[2][]{\texttt{#2}}%
\else
\DeclareGraphicsExtensions{.mps,.ps,.eps}
\usepackage{hyperref}
\hypersetup{%
  pdftitle={Pair Approximation Models for Disease Spread},
  pdfauthor={%
		J\'er\^ome Benoit,
		Ana Nunes and
		Margarida Telo da Gama},
  pdfsubject={PACS: 02.50.-r, 87.23.Ge, 05.70.Ln},
  pdfkeywords={complex systems; epidemic models; phase transitions},
  pdfpagemode=None,
  pdfstartpage=1,
  pdfhighlight=/O,
  pdfview=FitH,
  pdfstartview=FitH,
  urlcolor=blue,
  }
\IfFileExists{thumbpdf.sty}{\usepackage{thumbpdf}}{}
\providecommand{\eprint}[2][]{\href{http://arxiv.org/abs/#2}{\texttt{#2}}}%
\fi
\IfFileExists{datetime.sty}{%
  \usepackage{datetime}
	\providecommand{\now}{\xxivtime\space\today}
	}{%
	\providecommand{\now}{\the\time,\space\today}
	}

\input{\jobname.rty}

\begin{document}
\title{Pair Approximation Models for Disease Spread}
\author{%
	J\'er\^ome Benoit\thanks{\email{benoit@cii.fc.ul.pt}}
	\and
	Ana Nunes
	\and
	Margarida Telo da Gama
	}
\institute{%
	Centro de F\'isica Te\'orica e
		Computacional e Departamento de F\'\i sica,\\
	Faculdade de Ci\^encias da Universidade de Lisboa,\\
	Avenida Professor Gama Pinto 2,
	P-1649-003 Lisboa,
	Portugal%
	}
\date{\eprint{q-bio/0510005}}
\abstract{%
	We consider a Susceptible-Infective-Recovered (\textsc{SIR}) model,
	where the mechanism for the renewal of susceptibles is demographic,
	on a ring with next nearest neighbour interactions,
	and a family of correlated pair approximations (\textsc{CPA}),
	parametrized by a measure of the relative contributions of loops and open 
	triplets of the sites involved in the infection process.                   
	We have found that the phase diagram of the \textsc{CPA},
	at fixed coordination number,
	changes qualitatively as the relative weight of the loops increases,
	from the phase diagram of the uncorrelated pair approximation
	to phase diagrams typical of one-dimensional systems.
	In addition,
	we have performed computer simulations of the same model and shown that
	while the \textsc{CPA} with a constant correlation parameter
	cannot describe the global behaviour of the model,
	a reasonable description of the endemic equilibria
	as well as of the phase diagram may be obtained
	by allowing the parameter to depend on the demographic rate.
	}
\maketitle

\section{Introduction}\label{sec/introduction}
Stochastic Susceptible-Infective-Recovered (\textsc{SIR}) epidemic models
on lattices and networks can be mapped on to percolation problems
and are well understood \cite{Grassberger1982,MooreNewman2000May,MooreNewman2000Nov}.
To describe disease spread and persistence in a community,
the model must be extended to include a mechanism for renewal of susceptibles,
either births or immunity waning. 

Models with immunity waning,
Susceptible-Infective-Recovered-Susceptible (\textsc{SIRS}),
are based on the following transitions:
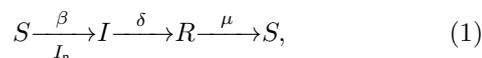
\begin{equation}
\label{ESWN/SIRS/scheme}
	{S}{\xrightarrow[I_{\eswnNeighbourIndex}]{\mspace{15.0mu}\eswnInfectionRate\mspace{15.0mu}}}%
	{I}{\xrightarrow{\mspace{15.0mu}\eswnRecoveryRate\mspace{15.0mu}}}%
	{R}{\xrightarrow{\mspace{15.0mu}\eswnDemographicRate\mspace{15.0mu}}}%
	{S},
\end{equation}
meaning that
any susceptible individual $S$ can be infected
by an infected neighbour $I_{\eswnNeighbourIndex}$
at the infection rate $\eswnInfectionRate$,
any infected individual $I$ becomes recovered $R$
at the recovery rate $\eswnRecoveryRate$,
and 
any recovered individual $R$ becomes susceptible $S$
at the immunity loss rate $\eswnDemographicRate$.
Following customary habits,
we shall choose time units
for which \mbox{$\eswnRecoveryRate\!=\!1$}.

The \textsc{SIRS} model interpolates between two well known models, 
the contact process
(also known as Susceptible-Infective-Susceptible or \textsc{SIS})
and the \textsc{SIR} model,
in the limits $\eswnDemographicRate\rightarrow\infty$ and $\eswnDemographicRate\rightarrow{0}$, respectively,
and much is known about its behaviour on regular lattices,
both from the point of view of rigorous results
\cite{DurrettNeuhauser1991,AndjelSchinazi1996,vdBerg1998}
and of assessing the performance of mean field
and pair approximations against stochastic simulations \cite{JooLebowitz2004}.

In particular, it is known \cite{vdBerg1998} that
on hypercubic lattices of arbitrary dimension
the phase diagram of \eqref{ESWN/SIRS/scheme} has two critical values,
\mbox{$\eswnCriticalInfectionRate(\infty)<\eswnCriticalInfectionRate(0)$},
which are the critical rates of the two limit problems
that is the contact process and \textsc{SIR}, respectively.
For \mbox{$\eswnInfectionRate<\eswnCriticalInfectionRate(\infty)$}
there is disease extinction for every $\eswnDemographicRate$,
while for $\eswnCriticalInfectionRate(0)<\eswnInfectionRate$
there is disease persistence for every $\eswnDemographicRate$.
For $\eswnCriticalInfectionRate(\infty)<\eswnInfectionRate<\eswnCriticalInfectionRate(0)$
disease persistence occurs only for $\eswnDemographicRate$ above a certain threshold.
The region of disease persistence for every $\eswnDemographicRate$ is `missing'
in dimension \mbox{$\eswnDimension\!=\!1$},
because in this case $\eswnCriticalInfectionRate(0)$ is infinite.

In \cite{JooLebowitz2004} the uncorrelated pair approximation 
(\textsc{UPA}, see Section~\ref{sec/MFAUPA})
was applied to the \textsc{SIRS} model \eqref{ESWN/SIRS/scheme} on linear and square lattices,
and the phase diagrams computed from the corresponding equations of evolution were compared
with the mean field phase diagram and with the results of simulations.
It was shown that,
by contrast with the mean field approximation,
the \textsc{UPA} phase diagram agrees qualitatively
with the simulations and the exact results both
in \mbox{$\eswnDimension\!=\!1$} and in \mbox{$\eswnDimension\!=\!2$}.

Since the \textsc{UPA} does not take into account the lattice dimensionality explicitly,
it predicts identical phase diagrams on lattices with the 
same coordination number $\eswnDegree$,
namely on linear (\mbox{$\eswnDimension\!=\!1$})
and square (\mbox{$\eswnDimension\!=\!2$}) lattices,
when next nearest neighbours (\mbox{$\eswnDegree\!=\!4$}) are considered.
However,
in one dimension the critical infection rate \mbox{$\eswnCriticalInfectionRate(0)=\infty$},
and the critical line has an asymptote at $\eswnDemographicRate=0$, while 
in two dimensions the critical line crosses the $\eswnDemographicRate=0$ axis at a 
finite value of $\eswnCriticalInfectionRate$,
which is the result of the \textsc{UPA} for \mbox{$\eswnDegree\!=\!4$}.

The question is then
whether generalized pair approximations can account
for the dependence on dimensionality and,
in particular, 
whether they can describe phase diagrams
with different qualitative behaviours at fixed coordination numbers.

We have addressed this question, and more generally the problem of 
constructing suitable pair approximations (Section~\ref{sec/MFAUPA}),
in the context of a modification of model \eqref{ESWN/SIRS/scheme},
where the mechanism of renewal of susceptibles is demography,
rather than immunity waning.
This is the natural scenario in the epidemiology of diseases that
confer permanent immunity,
such as childhood infectious diseases \cite{AndersonMay,MurrayII}.
For this model infection obeys the same rules as in \eqref{ESWN/SIRS/scheme},
immunity is permanent and all individuals,
whatever their state,
are submitted to birth and death events at a rate $\eswnDemographicRate$.
The stochastic process,
which describes the dynamics of this system,
is governed by the transitions
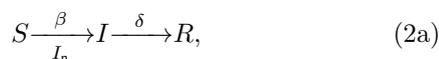
\begin{subequations}
\label{ESWN/scheme}
\begin{gather}
\label{ESWN/scheme/disease}
	{S}{\xrightarrow[I_{\eswnNeighbourIndex}]{\mspace{15.0mu}
\eswnInfectionRate\mspace{15.0mu}}}%
	{I}{\xrightarrow{\mspace{15.0mu}\eswnRecoveryRate\mspace{15.0mu}}}{R}%
	,
	\\
\label{ESWN/scheme/demographic}
	{\{S,I,R\}}{\xrightarrow{\mspace{15.0mu}\eswnDemographicRate\mspace{15.0mu}}}{S}
	.
\end{gather}
\end{subequations}
In the limit $\eswnDemographicRate=0$,
both models \eqref{ESWN/SIRS/scheme} and \eqref{ESWN/scheme}
coincide with \textsc{SIR} model.
In the opposite limit the dynamics of the two models are drastically different.
While,
in the limit $\eswnDemographicRate=\infty$,
\textsc{SIRS} coincides with the contact model \cite{JooLebowitz2004},
in the same limit the dynamics of \eqref{ESWN/scheme} is trivial:
it is driven by demography,
that keeps the entire population susceptible for any $\eswnCriticalInfectionRate$,
and thus $\eswnCriticalInfectionRate(\infty)=\infty$.  
We are interested in the regime,
where $\eswnDemographicRate$ is smaller than the recovery rate,
which is meaningful for the study of acute disease spread.
Although in this regime the dynamics is dominated by the infection and recovery processes
which are identical in both models,
the behaviour of \eqref{ESWN/scheme} appears to be different,
in a subtle way,
from that of \eqref{ESWN/SIRS/scheme} (Section~\ref{sec/MFAUPA}).

We have considered the demographic \textsc{SIR} model \eqref{ESWN/scheme}
on a linear lattice with periodic boundary conditions (ring)
and next nearest neighbour interactions, \mbox{$\eswnDegree\!=\!4$}.
We constructed a family of correlated pair approximations (\textsc{CPA}),
parametrized by $\eswnClosedFormParameter$,
a measure of the relative contributions of loops and open triplets of 
connected sites involved in the disease spread (Section~\ref{sec/CPA}).
For $\eswnClosedFormParameter=0$ the approximation reduces to the standard \textsc{UPA}
(Section~\ref{sec/MFAUPA}). 
The phase diagrams of the \textsc{CPA} show that
as $\eswnClosedFormParameter$ increases from $0$ to $\eswnRemarkableClosedFormParameter$
(see Section~\ref{sec/CPA}) 
the \textsc{CPA} interpolates between the \mbox{$\eswnDegree\!=\!4$} \textsc{UPA} critical behaviour
and the typical one-dimensional phase behaviour,
with $\eswnCriticalInfectionRate(0)=\infty$.
Finally,
we have simulated the demographic \textsc{SIR} model \eqref{ESWN/scheme}
on a ring, with \mbox{$\eswnDegree\!=\!4$}.
The results of the simulations indicate that
while the \textsc{CPA} with a constant value of $\eswnClosedFormParameter$ cannot describe
the global phase diagram of \eqref{ESWN/scheme}, 
a reasonable description of endemic equilibria
as well as of the phase diagram is obtained
when $\eswnClosedFormParameter$ is allowed to depend on the demographic rate $\eswnDemographicRate$
(Section~\ref{sec/CPA}).
This illustrates that in addition to describe
the dimensional crossover for lattices with coordination number \mbox{$\eswnDegree\!=\!4$}, 
the \textsc{CPA} can be made semi-quantitative providing an alternative to 
the stochastic simulations of individual based models.
We conclude in Section~\ref{sec/discussion} with a brief discussion of the results.

\begin{figure}[t]
	\begin{center}
		\includegraphics[width=1.0\linewidth]{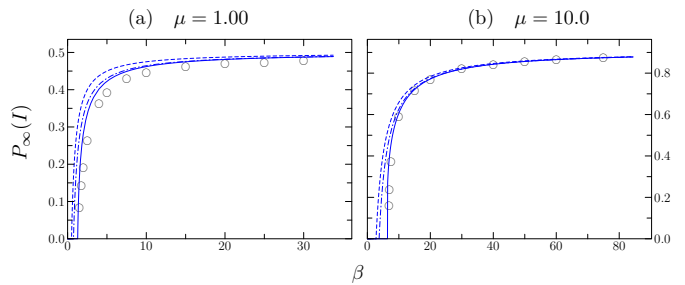}%
	\end{center}
	\caption{
		Endemic infective probability versus infection rate
				at (a) high demographic rate, and at (b) huge demographic rate:
			the endemic infective probability is plotted
			from simulations (open circles),
			the \textsc{MFA} (long dashed lines),
			the \textsc{UPA} (dashed dotted lines)
			and the correlated model with best-fit closed form parameters (solid lines).
			The fitting procedure is based on perpendicular offsets
			and on the assumption that the closed form parameter $\eswnClosedFormParameter$
			depends only on the demographic rate $\eswnDemographicRate$.
			Closed form parameters $\eswnClosedFormParameter$
			for (a) and (b) respectively:
			$0.50$,
			$0.70$.
		}%
\label{fig/X/endemic}
\end{figure}

\section{Mean Field and Uncorrelated Pair Approximations}\label{sec/MFAUPA}
In this section we consider the time evolution of the demographic \textsc{SIR} 
model on regular lattices and review the mean-field and (standard) uncorrelated pair approximations,
setting the notation and the stage for the development of the more sophisticated correlated pair approximations.

In the demographic \textsc{SIR} model on networks,
sites represent individuals and bonds social links.
The dynamics is governed by the stochastic process \eqref{ESWN/scheme}.
Denoting by $\prob{A}$ the probability for an individual to be in state $A$ (at time $t$),
$\prob{AB}$ the probability for a lattice bond to connect an individual in state $A$ to an individual in state $B$,
the time evolution of the singleton probabilities $\prob{A}$
can be described by the set of first order differential equations
\cite{AndersonMay,MurrayII}:
\begin{subequations}
\label{ESWN/SDE/1}
\begin{align}
\label{ESWN/SDE/1/S}
	\difft{\prob{S}}&=%
		+\eswnDemographicRate\:%
			\bigl[%
				\prob{I}%
				+%
				\prob{R}%
			\bigr]
		-\eswnInfectionRate\sum_{\eswnNeighbourIndex}{\prob{{S}{I_{\eswnNeighbourIndex}}}}
		,
		\\
\label{ESWN/SDE/1/I}
	\difft{\prob{I}}&=%
		+\eswnInfectionRate\sum_{\eswnNeighbourIndex}{\prob{{S}{I_{\eswnNeighbourIndex}}}}
		-(\eswnDemographicRate+\eswnRecoveryRate)\:%
			\prob{I}
		,
		\\
\label{ESWN/SDE/1/R}
	\difft{\prob{R}}&=%
		+\eswnRecoveryRate\:%
			\prob{I}
		-\eswnDemographicRate\:%
			\prob{R} 
		,
\end{align}
\end{subequations}
where the summations run over the connected neighbours.
Clearly the set of equations \eqref{ESWN/SDE/1} is not closed
since it involves pair probabilities without describing their time evolution.
This follows from the stochastic process \eqref{ESWN/scheme}
where infection \eqref{ESWN/scheme/disease} proceeds \textit{via} $SI$ contact pairs. 

As a matter of fact,
the time evolution of the $q$-tuple probabilities is described
by a set of first order differential equations expressing
their time derivatives as linear combinations of $q$-tuple and $(q\!+\!1)$-tuple probabilities,
subject to a normalization condition.
In order to proceed,
the set of equations must be closed,
that is the $(q\!+\!1)$-tuple probabilities must be written in terms of $q$-tuple probabilities.
The `art' is to use closures that capture
key physical features of the system and are still manageable by symbolic or numerical-symbolic computation.
The results of a particular closure, or approximation,
may then be checked against rigorous results and/or stochastic simulations.

For most closures the $(q\!+\!1)$-tuple probabilities are rational functions of the $q$-tuple probabilities,
appropriately normalized,
and thus the constrained set of first order differential equations may be replaced by an unconstrained set
where the time derivatives of independent $q$-tuple probabilities are expressed as
rational functions of these $q$-tuple probabilities.
Although the resulting sets of equations
are easily integrable by classical numerical methods
and admit polynomial systems as steady state equations,
their analysis remains cumbersome even at low order $q$.

\begin{figure}[b]
	\begin{center}
		\includegraphics[width=1.0\linewidth]{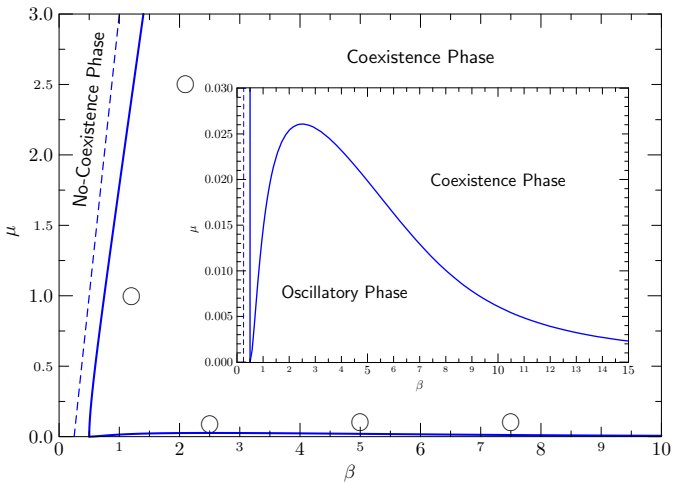}%
	\end{center}
	\caption{
		Phase diagram for the \textsc{UPA}:
			the no-coexistence phase and the coexistence phase
			are separated by
			the critical curve from 
			simulations (open circles),
			the \textsc{MFA} (long dashed line),
			and the \textsc{UPA} (thick solid line).
			Within the coexistence phase,
			at very low demographic rates
			$\eswnDemographicRate$,
			the \textsc{UPA} predicts
			an oscillatory phase
			as shown in the inset.
		}%
\label{fig/UPA/phasediagram}
\end{figure}

The simplest closure is the mean field approximation (\textsc{MFA}),
where the pairs ($2$-tuples) are assumed to be formed by
uncorrelated singletons ($1$-tuples):
\begin{equation}
\label{ESWN/UCS/approximation}
	\sum_{\eswnNeighbourIndex}{\prob{{S}{I_{\eswnNeighbourIndex}}}}\approx%
		\eswnDegree\:%
		\prob{S}\prob{I}
	.
\end{equation}
For the demographic \textsc{SIR} model the endemic equilibrium
(steady state)
is computed easily.
The mean-field endemic infective probabilities are plotted in Figure~\ref{fig/X/endemic}
as a function of the infection rate, at two values of $\eswnDemographicRate$.
For any value of $\eswnDemographicRate$,
the \textsc{MFA} predicts two different steady states:
at infection rates $\eswnInfectionRate$ smaller than
the critical infection rate $\eswnCriticalInfectionRate$
there is disease extinction,
while at infection rates $\eswnInfectionRate$ greater than
the critical infection rate $\eswnCriticalInfectionRate$ 
there is disease persistence,
\textit{i.e.} infected (and recovered) individuals coexist with susceptibles. 
The two regimes are separated by the mean-field endemic threshold 
that is plotted in Figure~\ref{fig/UPA/phasediagram} (dashed line).

We anticipate that the results of the \textsc{MFA} will be accurate
when the demographic process of \eqref{ESWN/scheme} dominates over the infectious one
since in this regime pairs are continually broken
and thus the behaviour of each individual is essentially independent on that of the other ones.
The infection process governed by \textsl{Susceptible}-\textsl{Infective} contact pairs,
dominates in the opposite regime (\mbox{$\eswnDemographicRate\ll{1}$}),
relevant in the epidemiological context.
The appropriate mean field theory is then the uncorrelated pair approximation (\textsc{UPA}).

The \textsc{UPA} is for pairs what the \textsc{MFA} is for singletons.
In the \textsc{UPA} triplets ($3$-tuples) are assumed to be formed
by uncorrelated pairs:
\begin{equation}
\label{ESWN/UCP/approximation}
	\sum_{\eswnNeighbourIndex}{\prob{{AS}{I_{\eswnNeighbourIndex}}}}\approx%
		(\eswnDegree-1)\:%
			\frac{\prob{SA}\prob{SI}}{\prob{S}}
	.
\end{equation}
The \textsc{UPA} is expected to outperform the \textsc{MFA} but,
in general,
its solution is not known in closed form.
For the demographic \textsc{SIR} model
the calculation of the phase diagram and the stability analysis
is still tractable by symbolic computation.
For lattices with coordination number \mbox{$\eswnDegree\!=\!4$}, the 
phase diagram is plotted in Figure~\ref{fig/UPA/phasediagram}.
It is clear that the \textsc{UPA} is quantitatively superior to the \textsc{MFA}
when compared with the results of simulations (open circles).
Both the \textsc{MF} and the \textsc{UP} approximations 
of the \mbox{$\eswnDegree\!=\!4$} demographic \textsc{SIR} model predict
a finite critical infection rate at $\eswnDemographicRate=0$,
while the simulations indicate
that $\eswnCriticalInfectionRate$ will diverge as $\eswnDemographicRate$ tends to $0$.

However,
the \textsc{SIRS} and the demographic \textsc{SIR} models
are different at low
(but finite)
demographic rates $\eswnDemographicRate$.
In the demographic \textsc{SIR} model the mechanism
for the renewal of susceptibles is totally random by contrast to
the mechanism of the \textsc{SIRS} model.
In our model susceptibles are born anywhere on the lattice while
in the \textsc{SIRS} model only previously infected sites loose immunity.

We note that
the randomizing effect of the demographic \textsc{SIR} mechanism
for the renewal of susceptibles is reminiscent
of the randomizing effect of shortcuts in small-world networks of the 
Watts and Strogatz type \cite{WattsStrogatz1998,Hastings2003}
where correlations are destroyed
and an effective mixing of the population is achieved,
with (drastic) consequences on the phase diagram. 

Finally, it is worth noticing that the \textsc{UPA} also predicts the existence 
of an oscillatory phase within the survival or coexistence phase
(\textit{i.e.}, to the right of $\eswnCriticalInfectionRate(0)$),
for small values of $\eswnDemographicRate$
(Figure~\ref{fig/UPA/phasediagram}).
The same is true for the \textsc{UPA} of process \eqref{ESWN/SIRS/scheme}
on the square lattice.
This behaviour will be difficult to identify in stochastic simulations,
since it may be blurred by large fluctuations and stochastic extinctions.

\begin{figure}[t]
	\begin{center}
		\includegraphics[width=1.0\linewidth]{pamds-PhaseDiagram-CPA}%
	\end{center}
	\caption{
		Isoparametric phase diagrams for the correlated model:
			the no-coexistence and coexistence phases
			are separated by the critical curve from
			simulations (open circles),
			the \textsc{MFA} (long dashed line),
			the \textsc{UPA}
			(dashed dotted line)
			and the correlated model for different
			$\eswnClosedFormParameter$
			(solid lines).
			For
			$\eswnRemarkableClosedFormParameter\!\approx\!{0.3807}$
			(bold solid line)
			the critical infection rate $\eswnCriticalInfectionRate$ tends asymptotically to infinity
			when the demographic rate $\eswnDemographicRate$ vanishes.
			Closed form parameters $\eswnClosedFormParameter$
			from left to right:
			$\tfrac{1}{4}$,
			$\eswnClosedFormParameter^{*}$,
			$\tfrac{1}{2}$,
			$\tfrac{5}{8}$,
			$\tfrac{3}{4}$.
		}%
\label{fig/CLM/phasediagram/isoparametric}
\end{figure}

\section{Correlated pair approximations}\label{sec/CPA}
In order to construct more realistic pair approximations,
we have investigated closure procedures inspired
by the geometrical structure of the lattice.

Within this perspective and as far as social triplets are concerned,
the ring of degree \mbox{$\eswnDegree\!=\!4$}
and the triangular lattice \mbox{($\eswnDegree\!=\!6$)}
are propitious networks
since their nearest-neighbour triplets split into two distinct classes:
`chain-like' (open) and `loop-like' (closed) triplets.
A very naive idea is to take into account the two classes of triplets
and to use the probability $\eswnClosedFormParameter$ and $1-\eswnClosedFormParameter$
of finding respectively a `loop-like' triplet and a `chain-like' triplet
as a parameter to be fitted to simulation results.
Thus triplets are assumed to be formed
either of uncorrelated (chained) pairs
or of correlated (looped) pairs
\cite{VanBaalen2000}:
\begin{equation}
\label{ESWN/CPL/approximation}
\begin{split}
	&\qquad%
	\sum_{\eswnNeighbourIndex}{\prob{{AS}{I_{\eswnNeighbourIndex}}}}\approx%
	\\
	&%
		\begin{cases}
			(\eswnDegree-1)\,%
				\Bigr[%
					\bigr(1-\eswnClosedFormParameter\bigl)\,%
					\tfrac{\prob{SA}\prob{SI}}{\prob{S}}%
					+%
					\,%
					\eswnClosedFormParameter\:%
					\tfrac{\prob{AI}\prob{SA}\prob{SI}}{\prob{A}\prob{S}\prob{I}}%
				\Bigl]%
				&\\
			\qquad\text{if ${A}\in\{S,R\}$}%
			,
			&\\
			(\eswnDegree-1)\,%
				\prob{SI}%
				-%
				\sum_{\eswnNeighbourIndex}{%
					\bigr[%
						\prob{{SS}{I_{\eswnNeighbourIndex}}}%
						\!+\!%
						\prob{{RS}{I_{\eswnNeighbourIndex}}}%
					\bigl]%
					}
			&\\
			\qquad\text{if ${A}={I}$}%
			.
			&
		\end{cases}
\end{split}
\end{equation}

The demographic \textsc{SIR} version of the \textsc{CPA} 
\eqref{ESWN/CPL/approximation}
is amenable by cumbersome numerical-symbolic computation although some 
interesting results may be obtained by symbolic computation.
The phase diagrams
are shown in Figure~\ref{fig/CLM/phasediagram/isoparametric}.
We find that, as $\eswnClosedFormParameter$ increases
from $0$ to $\eswnRemarkableClosedFormParameter\!\approx\!{0.3807}$%
\footnote{%
	the real solution of the cubic equation
	$27\eswnClosedFormParameter^{3}-18\eswnClosedFormParameter^{2}+87\eswnClosedFormParameter-32=0$.
	}%
,
keeping \mbox{$\eswnDegree\!=\!4$} fixed,
the \textsc{CPA} phase diagrams interpolate
between the \textsc{UPA} behaviour
and typical one-dimensional phase diagrams with $\eswnCriticalInfectionRate(0)=\infty$.
At $\eswnRemarkableClosedFormParameter$
the critical infection rate $\eswnCriticalInfectionRate$ tends asymptotically
to infinity as the demographic rate $\eswnDemographicRate$ vanishes.
Inspection of Figure~\ref{fig/CLM/phasediagram/isoparametric} also shows that
the closed form parameter $\eswnClosedFormParameter$ cannot be constant
if a quantitative description of the global phase diagram is required.
If we allow $\eswnClosedFormParameter$ to depend on $\eswnDemographicRate$,
reasonable descriptions of the endemic equilibria
(Figure~\ref{fig/X/endemic})
and of the global phase diagram
(Figure~\ref{fig/CLM/phasediagram/isoparametric})
are obtained.
For the \textsc{SIRS} model \eqref{ESWN/SIRS/scheme} on the square lattice,
a \textsc{CPA} obtained by fitting $\eswnClosedFormParameter$ to $\eswnCriticalInfectionRate(0)$
will improve the results of the \textsc{UPA} used
in \cite{JooLebowitz2004}
to describe the behaviour of the system at low values of $\eswnDemographicRate$.

\section{Discussion}\label{sec/discussion}
We have proposed a simple \textsc{CPA} that
was shown to provide a reasonable approximation
to the behaviour of stochastic models that are relevant in epidemiology
---
the agreement against simulation data
being far better than \textsc{MFA} and \textsc{UPA}
with a suitable choice of the parameters.
The resulting equations of evolution may be used to approximate phase diagrams,
as well as steady state and dynamical behaviours of the associated stochastic models.
The \textsc{CPA} takes into account some of the effects of the local lattice structure
and yields a clear alternative to heavy stochastic simulations.

One of the directions of future work includes the development of \textsc{CPA}s, 
along the lines of the present work,  
to account for the local (lattice like) structure of a class of 
complex networks,
such as the Watts and Strogatz small-world networks,
that have been shown to be relevant in epidemiological contexts
\cite{Verdasca2005}.

\begin{acknowledgement}
Financial supports from
the Foundation of the University of Lisbon,
under contract
BPD-CI/01/04,
and from
the Portuguese Foundation for Science and
Technology (FCT),
under contracts
POCTI/ESP/44511/2002
and
POCTI/ISFL/2/618,
are gratefully acknowledged.
\end{acknowledgement}

\bibliographystyle{apsrev}
\bibliography{pamds}

\end{document}